\documentclass[preprint2,epsf]{aastex}
\shorttitle{Absolute Proper Motion of the Fornax Dwarf}

\begin{document}

\title{Absolute Proper Motion of the Fornax Dwarf Spheroidal Galaxy
from Photographic and $HST$ WFPC2 Data}

\author{Dana I. Dinescu\altaffilmark{1,2,4}, Brian A. Keeney\altaffilmark{3,4},
Steve R. Majewski\altaffilmark{4}, and Terrence M.
Girard\altaffilmark{1}}

\altaffiltext{1}{Astronomy Department, Yale University, P.O. Box 208101,
New Haven, CT 06520-8101 (dana@astro.yale.edu, girard@astro.yale.edu)}
\altaffiltext{2}{Astronomical Institute of the Romanian Academy, Str.
Cutitul de Argint 5, RO-75212, Bucharest 28, Romania}
\altaffiltext{3} {Center for Astrophysics and Space Astronomy,
Department of Astrophysical and Planetary Sciences, 
University of Colorado, Box 389, Boulder, CO 80309}
\altaffiltext{4}{Department of Astronomy, University of Virginia, 
Charlottesville, VA 22903-0818}
\begin{abstract}

We have measured the absolute proper motion of the 
Fornax dwarf spheroidal galaxy
from a combination of photographic plate material and $HST$ WFPC2 data that provide a time
baseline of up to 50 years. The extragalactic reference frame consists of
8 QSO images and  48 galaxies.
The absolute proper motion is
$\mu_{\alpha}~cos~\delta = 0.59 \pm 0.16$ mas yr$^{-1}$ and
$\mu_{\delta} = -0.15 \pm 0.16$ mas yr$^{-1}$. The corresponding
orbit  of Fornax is polar, with an eccentricity of 0.27, and 
a radial period of 4.5 Gyr.  Fornax's current location is near pericenter. 
The direction of the motion of Fornax 
supports the notion that Fornax belongs to the 
Fornax-LeoI-LeoII-Sculptor-Sextans stream 
as hypothesized by
Lynden-Bell (1976, 1982) and Majewski (1994).

According to our orbit determination, Fornax crossed the Magellanic plane
$\sim 190$ Myr ago, a time that coincides with the termination of 
the star-formation process in Fornax. We propose that ram-pressure
stripping due to the passage of Fornax through a gaseous
medium denser than the typical intragalactic medium left behind
from the LMC may have caused the end of star 
formation in Fornax. The excess, anomalous clouds within the South Galactic
pole region of the Magellanic Stream whose origin has long been debated
in the literature as constituents of either the Magellanic Stream or
of the extragalactic Sculptor group, are found to
lie along the orbit of Fornax. We speculate that these clouds
are stripped material from Fornax as the dwarf crossed the 
Magellanic Clouds' orbit.

\end{abstract}


\section{Introduction}

The Fornax dwarf spheroidal galaxy (dSph) is the second most 
luminous of the ten dSph satellites of the Milky Way, just after
Sagittarius dSph (Sgr) (e.g., Grebel, Gallagher \& Harbeck 2003,
hereafter GGH03, Majewski et al. 2003). Among the well-known Milky-Way satellites,
Sagittarius and Fornax  are the only 
 dSph that have their own system of globular clusters\footnote[1]
{The recently discovered structure toward the Galactic Anticenter
also known as the Monoceros/Canis Major system
apparently contains old open clusters
as well as globular clusters (e.g., Frinchaboy et al. 2004, Martin et al. 2004).
Since it is a system rather poorly characterized at the present time,
we chose not to include it in the present discussion.
Nevertheless, it is yet another example of a system able to 
form clusters.}
(Hodge 1961, Da Costa \& Armandroff 1995). Sgr
has at least five globular clusters associated with it 
(Pal 12 besides the four clusters in the main body of Sgr, Dinescu et al.
 2000,
Mart\'{i}nez-Delgado et al. 2002, Cohen 2004), 
and arguably even more scattered along its
orbit (Bellazzini et al. 2003, Majewski et al. 2004) as the dwarf is in the process of
tidal disruption. Fornax is known to have five globular clusters.
These clusters show a strong second-parameter effect
as recently confirmed by Mackey \& Gilmore (2003b), while this is
not the case for the known Sagittarius clusters (see Table 4
 in Dinescu et al. 2001). Also, the Fornax clusters are all metal poor
($-2.0 \le$ [Fe/H] $\le -1.5$)
and have rather similar masses, while those of Sgr span a large range in metallicity
($-2.0 \le$ [Fe/H] $\le -0.6$)
and mass (see Strader et al. 2003, Table 7 in Gilmore \& Mackey 2003a,
Table 4 in Dinescu et al. 2001).

Fornax itself has a wide metallicity range
($-2.0 \le$ [Fe/H] $\le -0.4$) with a mean of -0.9 (e.g., Pont et al.
 2004), a characteristic 
shared only by Sgr among the Milky Way dSph satellites.
Fornax's population consists of old, metal-poor stars, a majority 
of intermediate-age stars  ($\sim4$ Gyr), and young stars
formed between 2 Gyr and 200 Myr ago (Pont et al. 2004, 
Saviane et al. 2000, Stetson, Hesser \& Smecker-Hane 1998). 
One might expect H I to be detected in Fornax,
given its recent star formation. However, Young (1999) found
no evidence of H I at the galaxy center at a column density detection limit
of $4.6 \times 10^{18}$ cm$^{-2}$, nor at a limit of 
$7.9 \times 10^{18}$ cm$^{-2}$ at one
core radius (i.e., $M_{HI} < 0.7 \times 10^6$ M$_{\odot}$, GGH03).
If star formation in
Fornax has ceased only recently, then it is important to
determine the orbit of the satellite and understand whether its
interaction with the Galaxy may have been responsible for the abrupt 
cessation. A much-favored mechanism 
for removing gas from a dSph is ram-pressure stripping as the satellite 
moves through the intragalactic medium (e.g., GGH03). It has been suggested 
that the star formation histories of 
satellite galaxies may be correlated to their orbits and perigalactic passages
(e.g., GGH03 and references therein, Mayer et al. 2001).
The complex star formation history of Fornax makes 
it an interesting subject from this standpoint.

Lynden-Bell (1976, 1982), Kunkel \& Demers (1976) and Kunkel (1979) 
pointed out that the Galaxy's dSphs, as well as a number of its globular clusters,
seem to collect along distinct planar alignments and proposed that these alignments
reflect distinct {\it orbital} planes.
One of these orbital planes is the Magellanic 
plane which consists of the Magellanic Clouds, the H I gas trailing the 
Magellanic clouds (also known as the Magellanic Stream, MS hereafter), 
Ursa Minor, Draco, and Carina. The second plane 
includes Fornax, Leo I, Leo II and Sculptor. Majewski (1994) 
reinvestigated this issue and added Sextans and perhaps
Phoenix to the Fornax-Leo-Sculptor alignment, which is now known as 
the FL$^2$S$^2$ plane. To explain the alignments,
a common origin via the disruption of larger satellites orbiting 
in these planes 
is usually invoked, with the smaller dSphs forming as tidal condensations 
or broken-off fragments during the
dynamical interaction between our Galaxy and a massive satellite (Kroupa 1997), 
much like the tidal dwarfs seen in extragalactic contexts (e.g., Knierman et al. 2003).
The parent satellite has been proposed to be the Large Magellanic Cloud (LMC) 
for the Magellanic plane, and Fornax for the FL$^2$S$^2$ plane (Lynden-Bell 1982).
A real physical association of these spatial alignments
can be tested by ascertaining whether the member objects share a common angular
momentum, and this requires determining absolute proper
motions of the dSphs.

The issues listed above motivate the present study of the proper motion
of the Fornax system.
A collection of thirteen photographic plates from large aperture 
telescopes together with numerous
$HST$ frames taken in the field of Fornax collectively spanning a time baseline of up to
50 years now make possible a deep, high-precision proper-motion study.
During the course of our work with these data a preliminary 
absolute proper motion of Fornax using only $HST$ data was published by Piatek 
et al. (2002, hereafter P02).  Their earlier study, with data spanning an
epoch difference of up to 
two years, used QSO images as references for the determination of the
absolute proper motion.
Our largely independent measurement allows a comparison with this $HST$-only 
proper-motion determination.

The paper is organized as follows: we describe the observational
data and the astrometric reductions in Section 2, 
and the absolute proper-motion determination in Section 3.
In Section 4 we discuss the velocity and orbit of Fornax
in relation with its star-formation history and the FL$^2$S$^2$ plane.
We summarize our results in Section 5.

\section{The Data}

\subsection{Observational Material}

The absolute proper motion of the Fornax dwarf galaxy was determined 
from the measurement of thirteen photographic plates and from six 
$HST$ WFPC2 data sets taken at different pointings in
the field of the dwarf galaxy.
The various combinations of plate and $HST$ WFPC2
data coverage lead to epoch differences varying between 20 and 50 years for
different areas of the field.
The characteristics of the photographic plates are listed in Table 1,
while those of the $HST$ WFPC2 data sets are listed in Table 2.
The oldest plates were taken with the Hale 5m reflector
(scale = $11\farcs12$ mm$^{-1}$) in the early fifties.
One of these (PH805B) includes Fornax Cluster \# 3
(NGC 1049) and the other two include Cluster \# 4.  
Six plates were taken with the CTIO 4m reflector (scale = $18\farcs6$ mm$^{-1}$)
in the mid seventies.
Of these, we used only three plates in the astrometric reduction,
namely 276, 282 and 287.
The remaining
three were taken in the blue passband at large hour angles, and a preliminary
proper-motion reduction indicated that these plates have very large color
terms that could not be entirely modeled and removed from the proper motions.
These plates were however used in the calibration of the photographic 
photometry. 
Another seven plates were taken with 
the Las Campanas Du Pont 2.5m reflector (scale $10\farcs92$ mm$^{-1}$)
between 1977 and 2001. Because of the large angular sizes (1.5$\times$1.5 degrees
on 20$\times$20 inch plates) and long exposure times
for most of the latter plates, special care was taken to observe using two guide
probes placed on opposite sides of the plate field.  With two probes
the effects of differential atmospheric refraction could be observed
and partially compensated for by slowly rotating the plate during
the exposure.

The $HST$ WFPC2 data were taken between 1995 and 2000 for various
purposes: studying the Fornax clusters (e.g., Buonanno et al. 1999),
determining the absolute proper motion of the Fornax dwarf from $HST$ 
measurements only (P02), and as parallel observations.

In Figure 1 we show the distribution of the observational material
in the field of Fornax dSph. In this plot, the center of Fornax is at (0,0),
i.e., at
$\alpha = 2^h 40^m 04^s$, $\delta = -34\arcdeg 31\arcmin $, J2000,
(Walcher et al. 2003).
We have also indicated the five
globular clusters in Fornax, and the eleven QSOs detected by
Tinney, Da Costa \& Zinnecker (1997), and Tinney (1999) in the background of
the Fornax galaxy. 
The outermost square shows the boundary of the input catalog determined from the
scan of Du Pont plate CD0100.
The large Du Pont plates have
plate centers that vary by a few arcmin (Table 1). By using the same input
catalog, objects on the edges of the input catalog
may have been missed in some of the large Du Pont plates. 
At any rate, these objects were not included in the final proper-motion
solution. $HST$ fields, as well as the CTIO and Hale areas also also indicated.

\subsection{Measurements}

The photographic plates were scanned with the Yale PDS microdensitometer.
First, we have created input catalogs based on the entire digitization of
plates CD0100, 288, PH805 and PH863. The area coverage of measurable images 
is as follows: 1) all Du Pont plates cover 1.22$\times$1.22 degrees, except
CD3302, which is a smaller size plate than the rest at an 
area of 0.77$\times$0.77 degrees, 2) all CTIO plates cover 
0.70$\times$0.70 degrees each, 3) the Hale 5m
plate PH805 covers 0.32$\times$0.32 degrees, while Hale plates PH863 and PH275 cover 
0.25$\times$0.25 degrees. Although the area coverage for plate CD0100 
includes all of the plates listed in Table 1, we made full-area scans of one CTIO
plate and the Hale plates, because these are deeper than CD0100.
From these scans, preliminary
positions, object diameters, magnitudes and classification 
were determined with the FOCAS software (Valdes 1982, 1993). Then,
based on these input lists, the plates were scanned in a 
fine-raster, object-by-object mode. For the Du Pont 2.5m and
Hale 5m plates we have used a pixel size of 12.7 $\mu$m, while for the
CTIO 4m plates we have used a 10 $\mu$m pixel, because the plate scale is larger
for these latter plates. Image positions and instrumental magnitudes
for each plate were determined using the Yale Image Centering routines
(two-dimensional Gaussian fit; Lee \& van Altena 1983). Objects
from various input catalogs were then positionally cross-correlated to create
a master list. 

Plate solutions that include transformation of coordinates among
same-epoch plates give us an estimate of the centering uncertainty
per object and
plate. Thus, well-measured stars on the Hale 5m plates have a centering  
uncertainty of 3.0 $\mu$m (33 mas); on the CTIO plates, 
1.4 $\mu$m (26 mas); and
on the Du Pont 2.5m plates, 2.0 to 3.0 $\mu$m (22-33 mas). 
The 2001 Du Pont plate
is of exceptional quality due to the fine-grained emulsion and seeing
conditions. We estimate a centering uncertainty of 1.2 $\mu$m (13 mas) based on
a coordinate transformation into WFPC2 coordinates (see below).

The entire archived $HST$ data set in the field of Fornax, as retrieved in 2000,
consists of 156 WFPC2 images taken between 1995 and 2000. The $HST$ fields
were targeted on all Fornax clusters, on QSO J0240-3434A+B (Tinney et al.
 1997),
and on two parallel data fields. All of the Fornax fields had multiple
exposures in one or two filters except for one parallel data field.
We chose not to use
the $HST$ data on clusters \# 1 and \# 5 in this analysis, because these clusters
lie at the edge of our field (Fig. 1) where geometric systematics 
of the photographic plate material are severe.
 
The WFPC2 images were processed using IRAF routines.
First, they were corrected for the 34th row effect described
in Anderson \& King (1999), and then cleaned for cosmic rays using
the mask technique described in Yanny et al. (1994). One parallel data
field that had only one exposure was left uncleaned.
For each $HST$ pointing, all of the frames were aligned and median stacked
to create a high signal-to-noise image. Object detection was performed on this stacked image 
using the DAOFIND routine in IRAF. The list of objects detected was 
then used on each individual frame to determine their centroids
and magnitudes. The centroids were determined with the help of the PSF
algorithm in IRAF, for which the analytic part is chosen to be 
an elliptical Moffat function with $\beta = 1.5$. The positions thus obtained
were corrected for distortion using the coefficients determined
by Anderson \& King (2003). Two WFPC2 data sets partly overlap
 in the region of Fornax cluster 4, namely the 1995 set and the 
1999 set. A coordinate transformation for the common stars in these two data
sets yields a positional uncertainty of 0.06 WFC pixels per star, or
6 mas. This value is larger than the 0.02-pixel nominal WFPC2 
positional precision obtained
in studies that use an empirically-determined PSF in rich 
stellar fields and a wealth of dithered images (e. g.,
Anderson \& King 2000). However, the positional uncertainty of the $HST$ data
is significantly better than the centering uncertainty on the photographic
plates. Thus, the random proper-motion uncertainty is dominated by the 
centering uncertainty on the photographic plates.
We have used only WFC data except for the field containing QSO J0240-3434A+B
which was centered on the PC. In the rest of the data sets, there are either
too few stars in the PC that can be matched with stars on 
the photographic plates, or there are too many stars when the PC is placed on a
cluster. In the latter case, the photographic images are hopelessly blended,
and no reliable cross-identification can be achieved.
From all of the $HST$ data sets, we were able to cross-identify
only a few extragalactic objects with the photographic plates. 
One is QSO J0240-3434A+B, a lensed quasar
with two distinct images separated by $\sim 6$ arcsec, and 
four relatively bright galaxies, 
all located in the parallel data set taken in 2000 (see Table 1). 

\section{Proper-Motion Measurement}

The reduction procedure is similar to that used 
in Dinescu et al. (2000 and 2001, hereafter D00 and D01). Since details are described 
thoroughly there, here we only briefly mention the necessary
steps involved in the reduction.    
All of the photographic-plate coordinates were first corrected for 
distortion using the coefficients determined by Chiu (1976) 
for the Hale plates, and by Cudworth \& Rees (1991) for 
the CTIO and the Du Pont plates.
Photographic $B$ and $V$ magnitudes were obtained by calibrating
instrumental magnitudes from the Du Pont and CTIO plates 
in the appropriate passband with Johnson $BV$ 
magnitudes from the CCD ground-based study of Saviane et al. (2000).
Typical magnitude uncertainties are $\sim 0.1$ mags.
The $BV$ colors thus obtained for all of the objects
are crucial for modeling color-dependent positional systematics 
known to affect the photographic astrometry.

Since the Du Pont 2.5m plates have the smallest distortion correction
among the three sets of plates, we have
selected plate CD2677 to serve as a master plate.
This plate has very good image quality and is taken close to the meridian,
thus minimizing color systematics in the positions due to 
differential color refraction. The remaining Du Pont plates 
were transformed into this master plate
using polynomials up to 6th order in each coordinate and, when needed, color terms.
The magnitude-dependent systematics
were modeled by using Fornax stars
selected from the color-magnitude diagram (CMD).
Fornax stars represent a system with a common motion while random 
Galactic field stars do not have this propertry.
Thus, using Fornax stars, one can separate the guiding-induced magnitude
systematic errors in the proper motions from the true, 
secular proper motion (for details see D00 and references therein). 
Since Fornax stars span a magnitude range from
$V = 18$ down to our plate limit, it is only up to this bright limit that 
the proper motions are assumed to be free of magnitude systematics.
A first iteration that included only the Du Pont
plates, 
in order to minimize systematics, produced preliminary 
relative proper motions with a time baseline of 24 years.
The stars that define the
transformation are well-measured and are predominantly Fornax stars.
These preliminary proper motions were used to update object positions 
for each CTIO and Hale plates which were subsequently introduced into the
proper-motion solution and modeled in the same manner as the Du Pont plates.
A second iteration that included all of the photographic plates
was done, and then the $HST$ WFPC2 data were introduced into the solution.
Each WFPC2 chip was transformed independently
using up to 2nd order terms, and between 10 and 50 reference
stars. No magnitude or
color-dependent systematics were detected in these transformations.
For the photographic plates,
between 3000 and 20,000 stars were used
in the plate solutions, depending on the area coverage and depth of the plate.
For each object, the proper motion is calculated from an unweighted linear
least-squares fit of positions as a function of time. The error in the proper
motion is given by the scatter about this best-fit line. Measurements 
that differ by more  than $0\farcs2$ from the best-fit line are 
excluded. Only objects that have at least three measurements and a 16-year
epoch difference were kept, and a relative proper motion and proper-motion
uncertainty calculated.
For the QSOs, the images on each plate were visually inspected, and
poor images were discarded from the fit.
The final proper motion catalog includes 38,059 objects.

Based on the FOCAS classification of objects on the digitized plate
CD0100, as well as from the plot of image peak as a function of image
radius, we selected a preliminary list of galaxies. These 
were then visually inspected on plate CD0100 to confirm their classification as
galaxies rather than blended stars.
The final list of galaxies
covering the entire area included 315 objects. Of these, only
268 galaxies have measurements that allow a
proper-motion determination in our catalog. Similarly, from the initial
list of 12 QSO images, only 11
have a proper-motion determination. The QSO that failed to have 
a proper-motion determination
is J0242-3424 (Tinney 1999), which is the easternmost quasar in
Fig. 1.

In Figure 2 we show the distribution of the proper-motion uncertainty
as a function of magnitude for stars, QSOs and galaxies, in 
each coordinate.
The median value for the proper-motion uncertainty per star, 
for well-measured stars
(i.e., $V = 18$ to 20), is between 0.8 and 1 mas yr$^{-1}$.
Galaxies have a median value of the proper-motion
uncertainty between 1.3 and 3 mas yr$^{-1}$ in the range $V = 17$ to 20.
The well-measured QSOs (i.e. $V = 19$ to 20.5 ) have between 
1 and 2 mas yr$^{-1}$ proper-motion error.

In Figure 3 we show
the relative proper motions for all objects in the catalog
as a function of magnitude, color and X, Y
coordinates. The top panels show the entire magnitude range, while the
rest of the panels include objects with magnitudes between $V = 18$ and 20.5.
The X coordinate is aligned with the right ascension with
X increasing eastward, and the Y coordinate is aligned with declination
with Y increasing toward north, at a scale of $10\farcs92$/mm.
Therefore, for this small area, $\mu_x$ is
practically identical to $\mu_{\alpha} cos \delta$, and
$\mu_y$ to $\mu_{\delta}$.
There are no significant trends in the proper motions
with either magnitude or color.
However, proper motions as a function of position on the plate 
show that geometric
systematics are left in the proper motions determined from the global
plate solution.
This is because of the limited ability to correct for distortion due to
the uncertainty in the position of the distortion center and 
the uncertainty in the distortion coefficients.

To correct for this, we assume that, locally, all proper motions are
systematically affected in the same way (see D00, D01). 
Thus, for each extragalactic
object we define a local reference system comprised of 30 stars.
The number of stars to be used in the local solution is determined
such that the average radius of the local system is smaller than 
the size over which systematics become important. In this case
the average radius of the local reference systems is 13 mm.  We have also
performed a solution based on 15 stars in the local reference system, and 
we found that this solution 
is indistinguishable from the 30-star solution.
The local reference stars are selected from the CMD to belong to the 
Fornax dwarf.
They are well-measured stars, i.e.,
 the proper-motion uncertainty in each coordinate
is $\le 2.0$ mas yr$^{-1}$, with at least six measurements, and
they lie in the magnitude range $V = 18$ to 21.
The proper motion of each extragalactic object obtained from the global
solution is corrected by subtracting the average proper motion of the
local reference system. This average is defined by the median value of the
proper motions of the 30 stars. In Figure 4 we show the proper motions of
galaxies and QSOs determined after the local correction has been applied,
as a function of $V$ magnitude and $B-V$ color. No significant trends
are detectable in these plots.

The absolute proper motion of the Fornax dwarf is determined from a 
selected sample of the extragalactic objects measured, as follows.
From the QSOs we keep only those with proper-motion uncertainty
$\le 2.0$ mas yr$^{-1}$ (see Fig. 2). We also discard object J0240-3437 from 
Tinney's (1999) list, since this is a large, bright ($V = 15.5$) galaxy.
We are thus left with eight images of seven QSOs. From the list of
galaxies we keep only objects with $V = 17$ to 20, with proper-motion
uncertainty $\le$ 3 mas yr$^{-1}$ in each coordinate, 
and that are spatially located within
the denser regions of the Fornax dwarf, such that the local reference system
is appropriately comprised of Fornax stars. The spatial extent of the
galaxies selected matches closely that of the QSOs as seen 
in Figure 5.  The proper-motion error limit for galaxies 
essentially assures that only well-centered galaxies are used.
Well-centered galaxies are those galaxies that have image shapes more closely
resembling those of stars.
Within this subsample of 54 galaxies, 6 have proper motions that are 
larger than 5 mas yr$^{-1}$ in one of the components. These are considered outliers
and are discarded from the final absolute proper-motion estimate;
we are thus left with 48 galaxies. 
As a check that our local solution represents an improvement over the
global solution we calculate the proper-motion dispersion for 
the 48 selected galaxies as given by the global solution and
by the local solution. We obtain for the global solution,
$\sigma_{\mu_x} = 1.63$ mas yr$^{-1}$ and, $\sigma_{\mu_y} = 1.68$ mas yr$^{-1}$,
while from the local solution we obtain $\sigma_{\mu_x} = 1.59$ mas yr$^{-1}$ and, 
$\sigma_{\mu_y} = 1.61$ mas yr$^{-1}$. Similarly, for the 8 QSO images we
obtain from the global solution $\sigma_{\mu_x} = 1.18$ mas yr$^{-1}$ and, 
$\sigma_{\mu_y} = 1.10$ mas yr$^{-1}$, while from the local solution
we obtain $\sigma_{\mu_x} = 0.89$ mas yr$^{-1}$ and, $\sigma_{\mu_y} = 0.76$ mas yr$^{-1}$.
In both cases the proper-motion dispersion decreases from the global to
the local solution, an indication that geometric systematics left
in the global-solution proper motions were
lessened from the local-solution proper motions.

We determine a weighted mean average for the absolute proper motion of the Fornax dwarf
from the selected QSOs and galaxies. The weights are given by the 
square of the proper-motion uncertainty of each extragalactic object.
The absolute proper motion of the Fornax dwarf galaxy with respect to
QSOs is $\mu_{\alpha} cos \delta = 0.28 \pm 0.30$ mas yr$^{-1}$, and
$\mu_{\delta} = -0.45 \pm 0.28$ mas yr$^{-1}$, and with respect to galaxies is
$\mu_{\alpha} cos \delta = 0.70 \pm 0.18$ mas yr$^{-1}$, and 
$\mu_{\delta} = -0.01 \pm 0.19$ mas yr$^{-1}$. 
The two determinations differ by 1.2$\sigma$ of their formal uncertainties.
In Figure 6 we show the proper-motion distribution
of the selected galaxies and QSOs, and their corresponding averages.
For the lensed quasar J0240-3434A+B that has two images within $6\arcsec$ of each other,
we obtain the following results:
1) image A has $\mu_{\alpha} cos \delta = 0.84 \pm 0.52$ and $\mu_{\delta} = -0.04 \pm 0.64$ 
mas yr$^{-1}$, with $V = 19.29$ and $B-V = 0.39$, 
and 2) image B has $\mu_{\alpha} cos \delta = 0.73 \pm 0.34$ and 
$\mu_{\delta} = -0.93 \pm 0.80$ mas yr$^{-1}$ with $V = 19.90$ and $B-V = -0.03$.
 While the proper motions agree within their formal uncertainties, we note that
the uncertainty in $\mu_{\delta}$ for the B component is rather large, but not
unusual for this magnitude (see Fig. 2).
The two QSO images have rather different magnitudes and colors, and any
residual systematics with these two quantities will be reflected in 
their proper motions.

The final absolute proper motion of Fornax is taken to be
the  error-weighted mean of the QSO and galaxy determinations.
This is: $\mu_{\alpha} cos \delta = 0.59 \pm 0.16$ mas yr$^{-1}$
and $\mu_{\delta} = -0.15 \pm 0.16$ mas yr$^{-1}$.

The recent absolute proper-motion determination of the Fornax dwarf
made by P02, which is based on $HST$ WFPC2 and STIS data alone
and uses 4 QSO images (i.e.,  the lensed QSO in one PC field, and two
QSOs in two STIS fields),
gives $\mu_{\alpha} cos \delta = 0.49 \pm 0.13$ mas yr$^{-1}$ and
$\mu_{\delta} = -0.59 \pm 0.13$ mas yr$^{-1}$. While the proper motion along
right ascension agrees with our value, that along declination is 
different from ours at the 2-$\sigma$ level. 
Left-over systematics are always invoked to explain the discrepancies between
different proper-motion determinations. To the best of our abilities
we have studied and hopefully eliminated proper-motions 
systematics with magnitude, colors and positions to a level
below our formal uncertainty, as the comparison of our results for galaxies and QSOs indicate.
The P02 result does agree within uncertainties with our QSO-based solution.
However, we note that the QSOs have colors significantly bluer than the majority 
of the Fornax stars used here and in the P02 study. These stars are basically
red giants, and the color range of galaxies better overlaps with that of Fornax
stars (see Figs. 3 and 4). Therefore we regard the result based on galaxies as extremely helpful
in gauging residual systematics. 

The QSOs in the P02 study are the brightest
(see their Fig. 9) and bluest objects on the $HST$ frames, and 
while trends with color and magnitude are not necessarily expected, these
were not investigated in the P02 study.
Another possible source of error in the P02 result is the
distortion correction. For the PC data, they have used the older 
distortion coefficients (Baggett et al. 2002) rather than the improved, 
more recent ones determined by Anderson \& King (2003) in the densely
populated field of $\omega$ Centauri. The fact that the orientation of the 
P02 frames was chosen to be
similar at different epochs would alleviate this problem to some extent.
There are however indications that the distortion is time-dependent as a
result of the change in the scale of the telescope known as the 
orbital ``breathing'' of the optical telesecope assembly (see Anderson \& King 2003).


\section{Discussion}

\subsection{Velocities and Orbits}

We calculate the velocity components of Fornax assuming
that the distance to Fornax is $138 \pm 8$ kpc, and the heliocentric radial
velocity is $53.0 \pm 3.0$ km s$^{-1}$ (Mateo 1998). 
We have adopted a standard solar motion of 
$(U_{\odot}, V_{\odot}, W_{\odot}) = (-10.0, 5.25, 7.17)$ km s$^{-1}$
(Dehnen \& Binney 1998) with respect to the Local
Standard of Rest (LSR). Here the $U$ component is positive outward from the
Galactic center, $V$ is positive toward Galactic rotation, and
$W$ is positive toward the North Galactic pole.
The adopted rotation velocity of the LSR is 
$\Theta_{0} = 220.0 $ km s$^{-1}$, and the solar circle radius is
8.0 kpc.  The uncertainties in the derived Fornax 
velocity components are determined based on
the uncertainties in the distance, radial velocity and proper motions.
We also present the proper motion
in the Galactic rest frame (i.e., with Solar and LSR
motion subtracted), in both equatorial and Galactic coordinates.
These proper motions and the velocity components are 
listed in Table 3. 
The first row for each object lists the proper motion and velocity
components for that object, while the second row lists the 
1-$\sigma$ uncertainty in our measurement of these quantities.
The velocity components are in a cylindrical coordinate system,
where $\Pi$ is positive outward from the Galactic center,
$\Theta$ is positive in the direction of Galactic rotation, and
$W$ is positive toward the North Galactic pole. We also list
the radial and tangential velocity components, where the radial
component is along the direction from the Galactic center to the
object. 
Along with these, we derive here the same quantities for
the Fornax proper-motion measurement of P02, and
for the Sculptor proper-motion measurement of Schweitzer et al. (1995).

In Figure 7 we show density contour plots of Fornax as derived from
the star counts of the raster scan of plate CD0100. The Fornax globular
clusters are also marked. Galactic rest frame proper motions
are indicated as follows.
Our proper-motion determination
is represented by the shorter arrow (points ENE), 
and its uncertainty is defined by the shaded circle.
The longer arrow to the SE represents the proper motion as determined by 
P02. Our proper-motion value
indicates a direction of motion closer to the direction of the major
axis of the dwarf, while that of P02 is practically
perpendicular to the major axis. Also, the size of our proper motion is
smaller than that of P02.
The position angle of our proper motion is 
$79\arcdeg \pm 26\arcdeg$, while that
of P02 is $145\arcdeg \pm 17\arcdeg$.
The Sculptor dSph is located at a position angle of $279\arcdeg$
from Fornax. The direction toward Sculptor is represented with a gray 
line in Fig. 7.
Thus, our proper-motion determination
indicates that Fornax is moving, within errors, along the direction
on the sky defined by Fornax and Sculptor, and away from Sculptor (Fig. 7).
The position angle of Sculptor's proper motion is $40\arcdeg \pm 27\arcdeg$
(Schweitzer et al. 1995), while Fornax lies at a position angle of
$99\arcdeg$ with respect to Sculptor.
Therefore Sculptor's direction of motion is some $2.2\sigma$ away from
the FL$^2$S$^2$ plane.

We calculate orbital elements by integrating orbits back in time over a period 
of 20 Gyr and using the Galactic potential model from
Johnston, Spergel \& Hernquist (1995). The integration time is chosen 
such that a few orbits are completed, and the orbital elements are averaged 
over the number of orbits. The uncertainties in the orbital elements
are determined from 300 integrations that have initial conditions
determined randomly from the uncertainties in the proper motions, radial 
velocity and distance (see Dinescu, Girard \& van Altena 1999, 
Dinescu et al. 2000).
In Table 4 we list the orbital parameters for the current 
Fornax absolute proper-motion determination and for that of P02, and for the Sculptor dwarf.
The total orbital energy should be regarded as being on a relative 
scale for comparison between measurements and should not be used as a bound
versus unbound criterion for Fornax, for example.
The orbital inclination is calculated as $90\arcdeg -$ sin$^{-1}
(L_z/L)$. 

Our proper-motion determination implies a low-eccentricity, 
polar orbit for the Fornax dwarf, with the current location at pericenter.
Similar integrals of motion for two or more objects indicate membership to
the same stream. Our measurement for Fornax gives
$E$, $L$ and $L_z$ that are larger than those of Sculptor. However,
the uncertainties in these quantities do not allow us to 
confidently rule out the membership of Sculptor to a stream that
includes Fornax, or the proposed FL$^2$S$^2$ stream.

Our estimate of the apocentric
radius of Fornax is close to the Galactocentric radii of
Leo I (250 kpc) and Leo II (205 kpc) (Mateo 1998), thus reinforcing
the hypothesis of the common origin of the FL$^2$S$^2$ stream.

The Sextans dSph with a current Galactocentric distance of
86 kpc (Mateo 1998), may qualify as a member of the FL$^2$S$^2$ stream. While
its location on the sky appears farther away from the orbital plane
of Fornax (see Figure 9 below) than those of Leo I and Leo II, 
an overlap of the orbits is within the
uncertainties associated with the Fornax orbit.

The Phoenix dSph is located at $\sim 450$ kpc (Mateo 1998)
from the Galactic center,
about twice the apocenter distance of Fornax's orbit, and is therefore a rather unlikely
member of any daughter stream of the Fornax dSph, unless substantial 
orbital decay occurred for the latter.

In contrast to these results, the P02 proper motion measurement for 
Fornax gives a rather eccentric 
orbit for Fornax, with the dSph more likely bound to the Local 
Group than to our Galaxy.  
Specifically, P02 argue against the membership of Fornax  to 
the  FL$^2$S$^2$ stream.

\subsection{The Orbit of Fornax and the Missing Gas}

The Fornax dSph has a population that consists of
predominantly intermediate-age ($\sim 4$ Gyr) stars, besides 
the old, metal-poor stars that are more traditional in dwarf
spheroidals (GGH03 and references therein). In addition to these
stars, Fornax has relatively young stars with ages between 
2 Gyr and 200 Myr (Pont et al. 2004 and references therein).
Stars in Fornax have a wide metallicity range, between [Fe/H] = -2.0
and -0.4 as recently demonstrated by  Pont et al. (2004).
More importantly,  Pont et al. (2004) determine an age-metallicity relationship
which implies that Fornax underwent an initial enrichment process
up to [Fe/H] = -1.0 about 3 Gyr ago, after which continued star formation that lasted
until some 200 Myr ago increased the metallicity to [Fe/H] $\sim -0.5$.
The inferred star-formation rate increased substantially over the last 
4 Gyr (Table 5 in Pont et al 2004). Given this star-formation history,
Fornax is expected to have some gas associated with it.
However, Young (1999) detected no H I gas within
one core radius of the galaxy. Specifically, there were no detections 
at the column density detection limits 
of $4.6 \times 10^{18}$ cm$^{-2}$
at the galaxy center and  $7.9 \times 10^{18}$ cm$^{-2}$ at one
core radius. Young (1999) mentions that there may 
be undetected H I 
between the core radius and the tidal radius of Fornax.
However, if the lack of gas is due to its removal by ram-pressure
stripping as Fornax moves through a homogeneous intragalactic medium,
then this gas is more easily stripped from the least dense regions
of the satellite, and these regions are more likely the outskirts rather 
than the center of the dSph (GGH03). 
Also, the bright, blue stars in Fornax associated with young,
main sequence stars (Stetson et al. 1998) are located in the central
region of the dSph, where there was no H I detection. 
Thus, as Pont et al. (2004) suggest,
it seems that Fornax is now at a particular moment in its
history, just after its ``gas death''. 

GGH03 explored whether ram-pressure stripping as Fornax moves through
a uniform, gaseous halo can explain the
lack of gas in Fornax. By using the velocity determined by P02 and 
adopting a gaseous halo density of $10^{-5}$ cm$^{-3}$, GGH03
estimate an upper limit for the Fornax gas density of $ 10^{-2}$ cm$^{-3}$
where ram-pressure stripping is efficient. The Fornax gas density 
obtained from this simple, pressure-balance estimation
is not that of an environment where stars 
could have formed recently, as is seen in Fornax. Therefore the lack of gas in Fornax
has to be explained by a more powerful gas-removal process than the simple one 
just described. Our lower value of the velocity of Fornax compared to
that of P02 (Table 3) would make even less efficient
the ram-pressure stripping caused by its motion through a uniform, low-density 
gaseous halo. The upper limit for the Fornax gas density stated
above corresponds to a column density of $\sim 10^{19}$ cm$^{-2}$, if one assumes a 
uniform density distribution
within one core radius. In this estimation, we have used a
core radius of $15\arcmin$ (Walcher et al. 2003), which 
corresponds to 0.6 kpc. 

From our orbit determination, there may be another 
plausible explanation for
the lack of gas in a system that was capable of producing stars some 
200 Myr ago. Fornax crossed the Magellanic plane 
relatively recently. Using the velocity and distance data from
van der Marel et al.  (2002) for the LMC, we have derived its orbit, and 
therefore the orientation of the Magellanic plane. According 
to our orbit calculations, Fornax crossed the Magellanic plane about $ 190$ Myr
ago, a recent event, considering the 4.5-Gyr orbital 
period of Fornax (Table 4). 
At this crossing, Fornax was located $\sim 147$ kpc from the Galactic
center, specifically at (X, Y, Z) = (5, -25, -145) kpc (the Sun is at (8.0, 0.0 0.0) kpc). 
In Figure 8 we show the orbits of the LMC (dark line) and of Fornax
(gray line) in the plane of the Galactic disk, and the X-Z plane
perpendicular to the Galactic disk. The orbits represent 10-Gyr integrations
backward in time. The current positions of the LMC and Fornax are indicated 
with filled symbols. We have also marked 200 and 500 Myr ago on the 
orbit of Fornax with cross symbols.
 While in the X-Z plot of the orbits, the LMC orbit
does not quite reach that of Fornax, moderate uncertainties in the orbits
due to proper-motion uncertainties easily allow for their actual 
crossing.

It is conceivable that gas from the LMC trails along 
its orbit beyond what is now known as the MS,
a 100$\arcdeg$ tidal filament of neutral hydrogen.
The recent N-body/SPH simulations of the LMC-Galaxy interaction 
by Mastropietro
et al. (2004) show that a large amount of gas ($\sim 10^{8}$ M$_{\odot}$)
 is lost along the orbit. These simulations include both tidal and
ram-pressure stripping.
Contrary to previous modeling of this interaction, Mastropietro
et al. (2004) find that the amount of stripped stars
is negligible compared to the amount of stripped gas.
The predictions are that the MS forms a great circle on the sky.  
Their Figure 2 shows that the lost gas is 
distributed in the plane of the orbit, and out to distances above and below
the Galactic plane of $\sim 140$ kpc. 
In fact, Braun \& Thilker (2004) report the detection of 
a diffuse northern ($\delta = 20\arcdeg-40\arcdeg$) 
extension of the MS, that has a column density
of $\sim 10^{17}$ cm$^{-2}$. According to the Mastropietro
et al. (2004) models, this detection corresponds to the distant, 
above-the-Galactic-plane part of the LMC orbit.
Thus Fornax may have undergone 
recent, efficient ram-pressure stripping by passing through 
the denser, inhomogeneous environment of the gas in the Magellanic plane,
along the orbit of the LMC. Whether gas from the LMC in this part of its
orbit (at Z = -140 kpc) has the appropriate density to cause
efficient ram-pressure stripping in Fornax, remains to be determined
from detailed models of the Clouds' interaction with the Galaxy.
Nevertheless, based on our calculated orbits, 
the crossing of the Magellanic plane took place at a time that 
coincides with the time when star formation ceased in Fornax.
To illustrate the geometry, in Figure 9
we show the projection on the sky of the entire (i.e., one period) orbit of
Fornax, and the most recent Gyr of the orbit of the LMC. 
The dark part of Fornax's orbit represents the most recent Gyr.
The current locations of Fornax,
the LMC, Sculptor, Leo I, Leo II, Sextans and Phoenix are indicated
with filled circles.
The MS is indicated as derived from the H I column density data
kindly provided to us by Mary Putman (Putman et al. 2003,
hereafter P03). In this 
plot, we highlight the density contour corresponding 
to N$_{HI} = 10^{19}$ cm$^{-2}$.

Besides mapping out the large-scale MS structure, P03 also analyze the 
spatial and velocity distributions of H I clouds in the MS
and in the direction of the Sculptor group, as cataloged 
in a previous paper (Putman et al. 2002).
The Sculptor group is an
association of galaxies at the South Galactic pole (SGP), and
at distances ranging from 2 to 4 Mpc. Roughly half of this group
falls on the same line of sight as a part of the MS, namely that section
in the vicinity of the SGP.
P03 (see also references therein) 
found that there are more clouds in this region than in any
other part of the MS. They also found that the distributions
of velocity and orientation of the elongated clouds with respect to the 
MS do not match those of the clouds in the bulk of the
MS (their Figures 11 and 14). 
In addition, they show that the velocity 
distribution of the clouds in this region of the MS
does not match that of the Sculptor group
galaxies either. Unlike distant, extragalactic clouds, these anomalous 
clouds show diffuse connections between one another,
and a likely velocity gradient
across the MS (see their Figures 7 and 13).
P03 conclude that these
``excess'' clouds (when compared to the MS cloud distribution) 
are more likely associated with some halo material rather than the
distant Sculptor group. P03 further note that the 
origin of the clouds may be linked to crossing tidal streams from Galactic satellites with
polar orbits.  This is exactly the type of orbit that we have calculated from
our proper-motion measurement of Fornax dSph.

In Figure 10 (a) (left panel) we show this region of the sky in a gnomonic projection with the 
SGP at the center of the projection. The orbits of Fornax and the LMC are shown. 
The Sculptor, Phoenix and Fornax dSphs are indicated, and 
the H I data from P03 are shown, with the 
N$_{HI} = 10^{19}$ cm$^{-2}$ contour highlighted.
This panel is to be compared with
the density map in P03 (their Figs. 5 and 12), and
with the velocity map in P03 (their Fig. 7). A reference feature 
common to all these figures that can be used to guide the eye, 
is the bifurcation of the MS at $b \sim -75\arcdeg$, and $l\sim 345\arcdeg$.
Figure 10(b) is a similar projection in which we show
the H I clouds cataloged by Putman et al. (2002) as individual sources.
From their catalog we have excluded sources associated with galaxies.
Each cloud is represented with a line segment that
indicates the orientation and size (major axis) of the cloud.
Only elongated clouds (i.e., minor/major axis $\le$ 0.7) that are small
(i.e., semi-major axis $\le 3\arcdeg$) are included in this plot.
The Putman et al. (2002) catalog is comprised of clouds with 
local-standard-of-rest velocities larger than 80 km s$^{-1}$ in absolute
value, a selection that aims to eliminate features associated with the Galactic
disk H I emission.
From Fig. 10 (b), one can see that there is an increase
in the number of clouds in the SGP region of the MS, specifically
near the location on the sky of the
Sculptor dSph and between $l \sim 0\arcdeg$ to $45\arcdeg$, and
$b \le -60\arcdeg$. These excess clouds, already mentioned in
P03, are located along the newly determined orbit of Fornax.
Moreover, the orientation of the elongation of these
excess clouds is aligned with the orbit of Fornax,
while the bulk of the clouds along the MS, are aligned 
with the long axis of the MS. 
The orientation of the elongation of these two groups of 
clouds with respect to the long axis of the MS can also be seen in Figure 8 of
P03. In addition to the orientation of the clouds, Figure 8 of
P03 also shows the head-tail structure of the clouds, i.e., 
a dense core with diffuse tail structure. P03 note that often
the clouds aligned with the MS have the tails point away from the 
Magellanic Clouds. From Figure 8 of P03, the clouds in the SGP/Sculptor-group
 region of the MS that have an orientation almost perpendicular to the long axis of the 
MS have the tails point away from the MS, and, according to
our Figure 10 (b), away from Fornax. Therefore the orientation of the elongation
and of the head-tail structure of the clouds is similar for these two groups of
clouds, provided that for one group the source where they originated 
is the LMC, and for the other group the source is the Fornax
dSph.

From our orbit calculation, Fornax and/or material from Fornax
has a Galactocentric radial velocity
(in the Galactic rest frame) of $\sim -50$ km s$^{-1}$, when it crosses
the Magellanic plane. This negative, rather low radial velocity
indicates that Fornax is approaching a turning point in its orbit --- in this case 
the pericenter --- and it is well
within the velocity range of the excess clouds discussed in P03.
(See their Fig. 11, in which the Galactic-rest-frame velocities
are along the line of sight
rather than along the direction to the Galactic center, but
these directions are very close for objects located at
large Galactocentric radii.)

Therefore we suggest that the excess clouds in the SGP/Sculptor-group
region of the MS are stripped material from the Fornax dSph, as it 
crossed the Magellanic orbit (Fig. 8).

According to our orbit, Fornax's previous pericenter passage was 
$\sim 4.5$ Gyr ago (Tab. 4). Since there is a significantly larger 
intermediate-age (4-7 Gyr, e.g.,  GGH03, Tolstoy et al. 2003) population 
than old population in Fornax, it is conceivable that the intermediate-age
population is the result of the tidal interaction between Fornax 
and our Galaxy at the previous pericenter passage of the satellite.
Between the last pericenter passage and the present time
(which practically coincides with pericenter passage), Fornax continues
to form stars and consumes most of its gas. 
The recent crossing of the Magellanic plane (and the supposed 
left-over gas from the LMC)
may have swept away the remaining gas, causing star formation to cease.

\subsection{Estimation of the Mass of the Galaxy}

If Fornax is gravitationally bound to the Galaxy, then its current
velocity is smaller than the escape velocity. By assuming a point-mass
Galactic gravitational potential, the limit on the escape velocity
gives a lower limit for the Galactic
mass as:
\begin{equation}
M = {\frac{R_{GC} V^2}{2G}}
\end{equation}
where $R_{GC}$ is the Galactocentric radius, and $V$ is the
total velocity of the dSph. By taking $R_{GC} = 140$ kpc, and the 
total velocity from the values in Table 3 we obtain
$M = (9.4 \pm 7.2) \times 10^{11}$ M$_{\odot}$. The uncertainty in the mass estimate
is derived from that in the total velocity.
This estimate is in reasonable agreement with recent determinations
of the Milky Way mass from a large sample of halo objects and satellites
(Sakamoto, Chiba \& Beers 2003). They find a lower mass limit of 
$2.2\times10^{12}$ M$_{\odot}$.

\section{Summary}
We have measured the absolute proper motion of Fornax dSph
from a combination of photographic plates and $HST$ WFPC2 images with a time baseline
of up to 50 years. A total of 8 QSO images and 48 galaxies were used in the
correction to absolute proper motion. The uncertainty in each
proper-motion direction is 0.16 mas yr$^{-1}$. This
measurement implies a polar orbit, with a low eccentricity, and with 
the current location of Fornax near pericenter. The motion of Fornax 
supports the notion that Fornax belongs to the plane defined by 
Leo I, Leo II, Sextans, Sculptor and Fornax. We also note that the orbit
of Fornax indicates that the dSph crossed the Magellanic plane some 200 Myr
ago, a time that coincides with the termination 
of the dSph's star-formation process. 
It is found that the excess, anomalous H I clouds in the SGP/Sculptor-group
region of the MS (P03),
are located along the orbit of Fornax. The orientation of the elongation
of these excess clouds seems to better align with the orbit of Fornax
rather than with that of the LMC.   
 We thus speculate
that these clouds were stripped from Fornax, as the dSph crossed
the Magellanic orbit.

We acknowledge funding from $HST$ archive grant 08739.
SRM, DID and BK were also supported by NSF CAREER Award AST-9702521, 
NSF grant AST-0307851, NASA/JPL contract 1228235,
a Cottrell Scholar Award from the Research Corporation, and a 
David and Lucile Packard Foundation Fellowship to SRM.  
SRM thanks
Eduardo Hardy for loan of his early Du Pont plates of Fornax, 
Allen Sandage for loan of the 200-inch Fornax plates, Oscar Duhalde
and Fernando Peralta for assistance during SRM's Du Pont photographic 
observing runs, and David Monet for generating early plate scans of 
the Du Pont Fornax plates.
DID thanks Mary Putman for making available the H I density data for the MS.

This research has made use of the $SIMBAD$ database, operated at $CDS$,
Strasbourg, France.

\newpage
\clearpage

\begin{table*}[tbh]
\begin{center}             
\begin{tabular}{lcrclcc}                                 
\multicolumn{7}{c}{\bf Table 1. Photographic Plate Material } \\ \\
 \hline\hline   \\
\multicolumn{1}{l}{Plate \#}&\multicolumn{1}{c}{Date} &
\multicolumn{1}{c}{H.A.}  &\multicolumn{1}{c}{Exp. Time}&      
\multicolumn{1}{c}{Emulsion+Filter} &\multicolumn{2}{c}{R.A. (J2000)  Dec.}\\ 
& \multicolumn{1}{c}{(dd.mm.yy)} & \multicolumn{1}{c}{(hours)} & \multicolumn{1}{c}{(minutes)} & 
& \multicolumn{1}{c}{($h~m~s$)} & \multicolumn{1}{c}{($\arcdeg~~\arcmin~~\arcsec$)} \\ 
\hline \\
\multicolumn{7}{c}{Hale 5 m ($11\farcs12$ mm$^{-1}$)} \\  \hline \\        
 PH275B & 09.10.50 & 0.15 & 60 & 103a-E GG11+No. 25 & 
 $2~40~03~$ & $-34~30~27$ \\
 PH805B & 04.09.53 & 23.58 & 30 & 103a-D GG11 & 
 $2~39~48~$ & $-34~15~20$ \\
 PH863B & 13.09.53 & 23.15 & 60 & 103a-D Corning 3484 &
 $2~39~58~$ & $-34~28~52$ \\ \hline \\
\multicolumn{7}{c}{CTIO 4 m ($18\farcs6$ mm$^{-1}$ )} \\  \hline \\        
 276 & 18.12.74 & 23.06 & 45 & IIIa-J GG385 &
 $2~39~10~$ & $-34~40~56$ \\
 $278^1$ & 18.12.74 & 20.34 & 60 & IIIa-J GG385 &
 $2~39~10~$ & $-34~40~56$ \\
 282 & 19.12.74 & 21.53 & 70 & IIa-D  GG495 &
 $2~39~10~$ & $-34~40~56$ \\
 $283^1$ & 19.12.74 & 20.17 & 50 & IIIa-J GG385 &
 $2~39~10~$ & $-34~40~56$ \\
 287 & 20.12.74 & 21.80 & 70 & IIa-D  GG495 &
 $2~39~10~$ & $-34~40~56$ \\
 $288^1$ & 20.12.74 & 20.30 & 60 & IIIa-J GG385 &
 $2~39~10~$ & $-34~40~56$ \\ \hline \\
\multicolumn{7}{c}{Las Campanas DuPont 2.5 m ($10\farcs92$ mm$^{-1}$)} \\  \hline \\             
 CD0100 & 12.10.77 & 22.43 & 90 & 103a-D W16 & 
 $2~39~39~$ & $-34~35~41$ \\
 CD0103 & 13.10.77 & 22.60 & 60 & 103a-D W16 & 
 $2~39~29~$ & $-34~35~41$ \\
 CD2644 & 16.08.85 & 23.75 & 60 & 103a-O GG385 & 
 $2~39~57~$ & $-34~31~27$ \\
 CD2677 & 20.08.85 & 23.45 & 60 & 103a-O W2C & 
 $2~39~57~$ & $-34~31~27$ \\
 CD3107 & 02.11.94 & 23.33 & 180 & IIa-D  GG495 & 
 $2~39~55~$ & $-34~31~26$ \\
 CD3110 & 03.11.94 &  0.07 & 225 & IIa-D  GG495 & 
 $2~39~55~$ & $-34~31~26$ \\
 CD3302 & 23.01.01 & 20.37 & 120 & IIIa-F GG495 & 
 $2~40~08~$ & $-34~34~20$ \\
\hline 
\multicolumn{7}{l}{$^1$ These plates were used in the
photometric calibration, but not in the astrometric reduction.} \\        
\end{tabular}                                                    
\end{center}                                                     
\end{table*}                 

\clearpage

\begin{table*}[tbh] 
\begin{center}             
\begin{tabular}{lcclcc}                                 
\multicolumn{6}{c}{\bf Table 2. $HST$ WFPC2 Data } \\ \\
 \hline\hline   \\
\multicolumn{1}{l}{Target}
&\multicolumn{1}{c}{Date} &\multicolumn{1}{c}{Exp. Time}&      
\multicolumn{1}{c}{Filter} &\multicolumn{2}{c}{R.A. (J2000)   Dec.}\\ 
& \multicolumn{1}{c}{(dd.mm.yy)} & \multicolumn{1}{c}{(seconds)} & 
& \multicolumn{1}{c}{($h~m~s$)} & \multicolumn{1}{c}{($\arcdeg~~\arcmin~~\arcsec$)} \\ 
\hline \\
 Cluster 4 & 10.03.95 & 1100 & F814W &
 $2~40~07.94$ & $-34~32~19.0$ \\
 Cluster 3[NGC 1049] & 04.06.96 & 120-700 & F555W/F814W &
 $2~39~49.37$ & $-34~15~05.5$ \\
 Cluster 2 & 05.06.96 & 120-700 & F555W/F814W &
 $2~38~45.52$ & $-34~48~06.5$ \\
 Fornax Field[Parallel Data] & 25.09.98 & 300 & F606W &
 $2~39~31.75$ & $-34~33~50.6$ \\
 QSO[A+B] \& Cluster 4 & 10.03.99 & 160 & F606W &
 $2~40~05.36$ & $-34~34~11.0$ \\ 
 Fornax Field[Parallel Data] & 08.03.00 & 400 & F606W  &
 $2~38~34.45$ & $-34~39~04.8$ \\
\hline
\end{tabular}                                                    
\end{center}                                                     
\end{table*}

\begin{table*}[tbh] 
\begin{center}             
\begin{tabular}{lrrrrrrrrr}                                 
\multicolumn{10}{c}{\bf Table 3. Galactic-Rest-Frame Proper Motions and Velocities} \\ \\
 \hline\hline   \\
\multicolumn{1}{l}{Object}&\multicolumn{1}{c}{$\mu_{\alpha}$ cos $\delta$} &
\multicolumn{1}{c}{$\mu_{\delta}$}  &\multicolumn{1}{c}{$\mu_l$ cos $b$}&      
\multicolumn{1}{c}{$\mu_b$} & \multicolumn{1}{c}{$\Pi$} &
\multicolumn{1}{c}{$\Theta$} & \multicolumn{1}{c}{$W$} &
\multicolumn{1}{c}{$V_r$} & \multicolumn{1}{c}{$V_t$} \\
& \multicolumn{2}{c}{(mas yr$^{-1}$)} & \multicolumn{2}{c}{(mas yr$^{-1}$)}
& \multicolumn{3}{c} {(km s$^{-1}$)} &
\multicolumn{2}{c} {(km s$^{-1}$)}\\ 
\hline \\
Fornax       &  0.36 & 0.07 & -0.13 & 0.34 &
  197 & 62 & 123 & -23 & 239 \\
          & $\pm0.16$     &  0.16    &   0.16    &   0.16   &
  101  &  102      &  44      &    60     &    93  \\ \\
Fornax (P02) &  0.26 & -0.37 & 0.32 & 0.32 &
  152 & -230 & 116 & -38 & 296 \\
          &  $\pm0.13$    &  0.13 &   0.13 &  0.13 &
   81   &   82     &   36  &   48  &  79  \\ \\
Sculptor &  0.36 & 0.43 & -0.47 & -0.30 &
  60 & 193 & -88 & 96 & 199 \\
         &  $\pm0.22$     &  0.25    &   0.22    &  0.25   &
 90   &    87     &   12     &   16     &  87 \\
\hline
\end{tabular}                                                    
\end{center}                                                     
\end{table*}

\begin{table*}[tbh]
\begin{center}    
\begin{tabular}{rrrrrrrrrrr}
\multicolumn{11}{c}{\bf Table 4. Orbital Elements} \\ \\
\hline
\hline
\\
\multicolumn{1}{c}{Object} & 
\multicolumn{1}{c}{$E$}&
\multicolumn{1}{c}{$L_{z}$}&
\multicolumn{1}{c}{$L$} &
\multicolumn{1}{c}{$P_{\varphi}$}& 
\multicolumn{1}{c}{$P_r$}&
\multicolumn{1}{c}{$R_{a}$}&
\multicolumn{1}{c}{$R_{p}$}&
\multicolumn{1}{c}{$z_{max}$} &
\multicolumn{1}{c}{e} & 
\multicolumn{1}{c}{$\Psi$} \\
& \multicolumn{1}{c}{($10^{4}$km$^{2}$s$^{-2}$)} &
\multicolumn{2}{c}{($10^{4}$kpc km s$^{-1}$)} &
\multicolumn{2}{c}{($10^{9}$ yr)}& 
\multicolumn{1}{c}{(kpc)} &
\multicolumn{1}{c}{(kpc)} &
\multicolumn{1}{c}{(kpc)} &&
\multicolumn{1}{c}{($\arcdeg$)} \\ \\
\hline \\ 
\multicolumn{1}{l}{Fornax } & \multicolumn{1}{r}{5.4} &
\multicolumn{1}{r}{0.4} & \multicolumn{1}{c}{3.4} &
\multicolumn{1}{r}{6.0} & \multicolumn{1}{r}{4.5} &
\multicolumn{1}{r}{239} &
\multicolumn{1}{r}{138} & \multicolumn{1}{r}{186} &
\multicolumn{1}{r}{0.27} & \multicolumn{1}{r}{83} \\
                            & \multicolumn{1}{r}{$\pm1.4$} &
\multicolumn{1}{r}{0.4} &\multicolumn{1}{c}{1.8}  &
\multicolumn{1}{r}{2.8} &  &
\multicolumn{1}{r}{139} &
\multicolumn{1}{r}{19} & \multicolumn{1}{r}{54} &
\multicolumn{1}{r}{0.16} &  \multicolumn{1}{r}{14} \\ \\
\multicolumn{1}{l}{Fornax (P02)} & \multicolumn{1}{r}{6.9} &
\multicolumn{1}{r}{-1.4} & \multicolumn{1}{c}{4.2} &
\multicolumn{1}{r}{10.0} & \multicolumn{1}{r}{7.0} &
\multicolumn{1}{r}{442} &
\multicolumn{1}{r}{138} & \multicolumn{1}{r}{230} &
\multicolumn{1}{r}{0.52} & \multicolumn{1}{r}{70} \\
                           & \multicolumn{1}{r}{$\pm1.2$} &
\multicolumn{1}{r}{0.3} &\multicolumn{1}{c}{1.3}  &
\multicolumn{1}{r}{2.4} &  &
\multicolumn{1}{r}{171} &
\multicolumn{1}{r}{7} & \multicolumn{1}{r}{68} &
\multicolumn{1}{r}{0.16} &  \multicolumn{1}{r}{12} \\ \\
\multicolumn{1}{l}{Sculptor } & \multicolumn{1}{r}{2.8} &
\multicolumn{1}{r}{0.2} & \multicolumn{1}{c}{1.6} &
\multicolumn{1}{r}{2.8} & \multicolumn{1}{r}{2.1} &
\multicolumn{1}{r}{120} &
\multicolumn{1}{r}{58} & \multicolumn{1}{r}{91} &
\multicolumn{1}{r}{0.35} & \multicolumn{1}{r}{83} \\
                             & \multicolumn{1}{r}{$\pm1.2$} &
\multicolumn{1}{r}{0.1} &\multicolumn{1}{c}{0.9}  &
\multicolumn{1}{r}{1.1} &   &
\multicolumn{1}{r}{51} &
\multicolumn{1}{r}{15} & \multicolumn{1}{r}{31} &
\multicolumn{1}{r}{0.10} &  \multicolumn{1}{r}{7} \\
\hline 
\end{tabular}
\end{center}
\end{table*}

\begin{figure*}
\epsscale{1.5}
\plotone{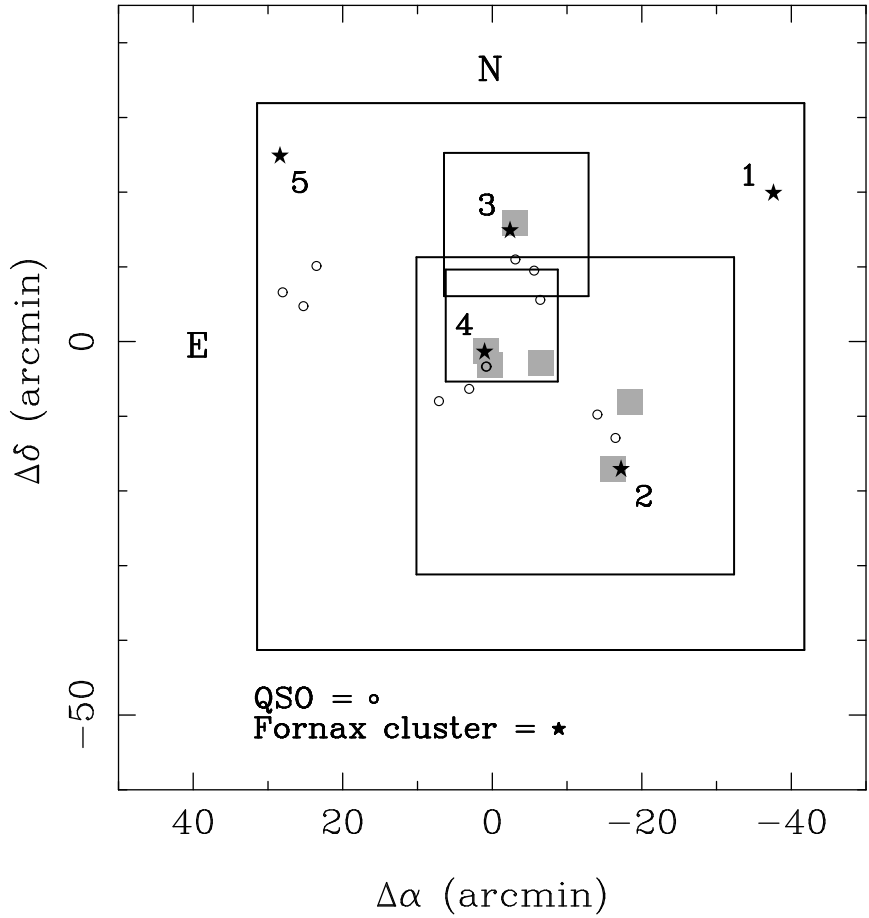}
\caption{Sky coverage of the observational material.
The outermost square shows the boundary of the area measured on the
large Du Pont plates.
The next largest square indicates the area covered by the CTIO plates.
The remaining two squares, containing clusters 3 and 4, show the
Hale area boundaries.  
The positions (but not the orientations)
of the $HST$ WFPC2 observations are indicated by the small shaded squares. 
The center of the Fornax dSph is at (0,0).
Fornax clusters are represented by star symbols and are labeled, and
the open circles indicate the positions of the QSO images.}
\end{figure*}

\clearpage

\begin{figure*}
\epsscale{1.8}
\plotone{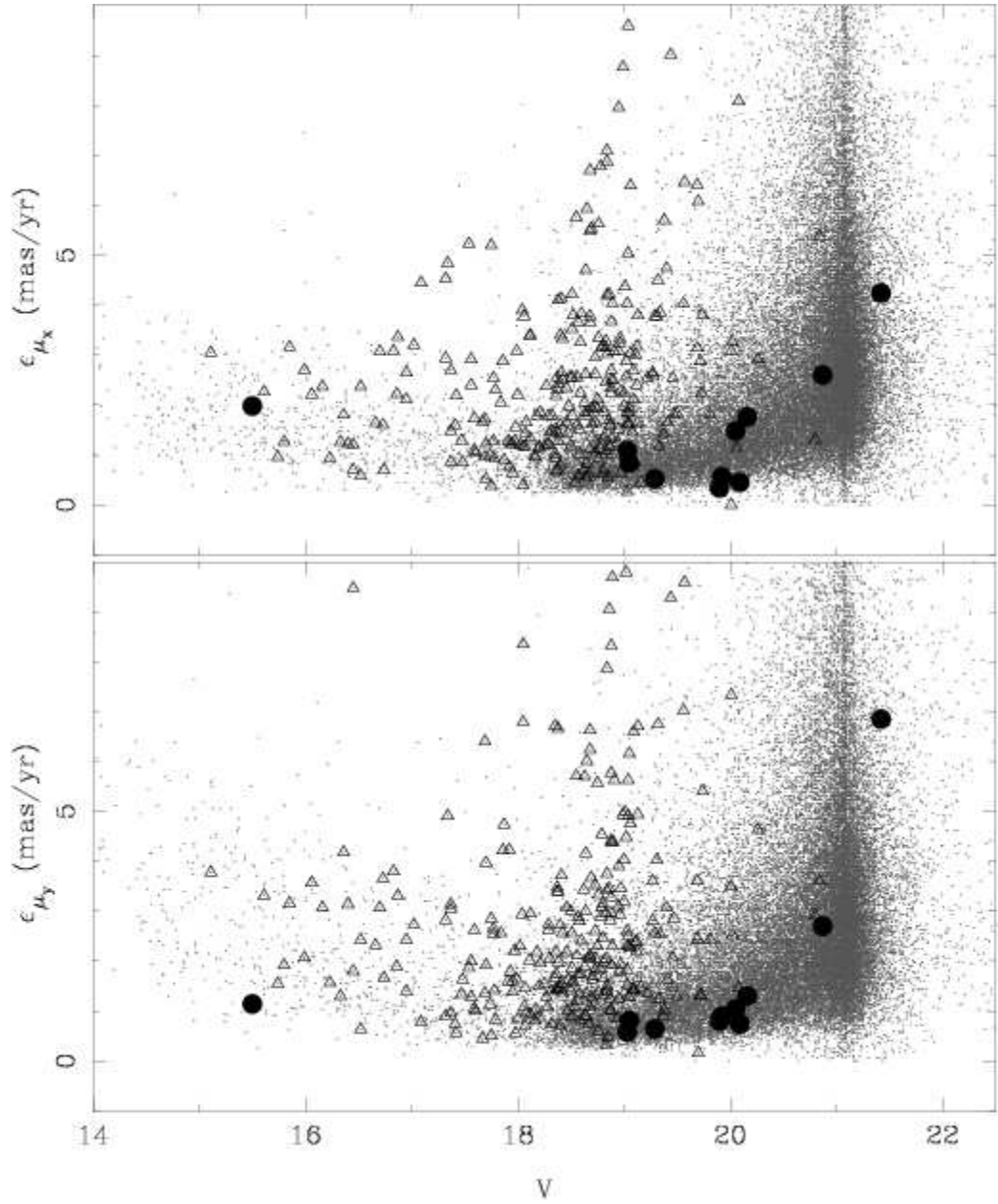}
\caption{Proper-motion errors as a function of magnitude. Filled circles
are the QSOs, open triangles are the galaxies, and dots are the stars.}
\end{figure*}

\begin{figure*}
\epsscale{2.0}
\plotone{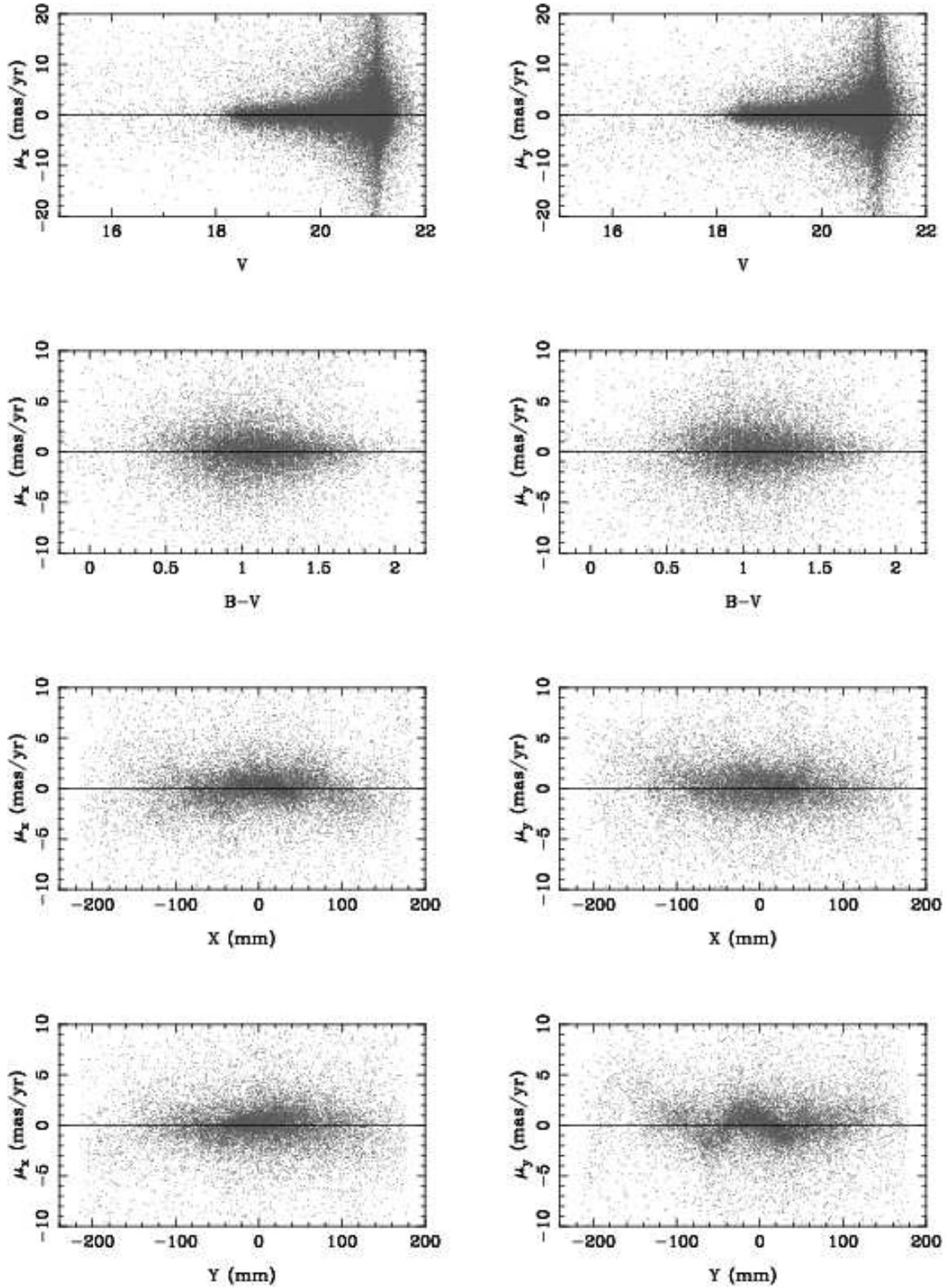}
\caption{Relative proper motions as a function of magnitude, color and
position of the plate for all objects in the catalog.}
\end{figure*}

\begin{figure*}
\includegraphics[angle=270]{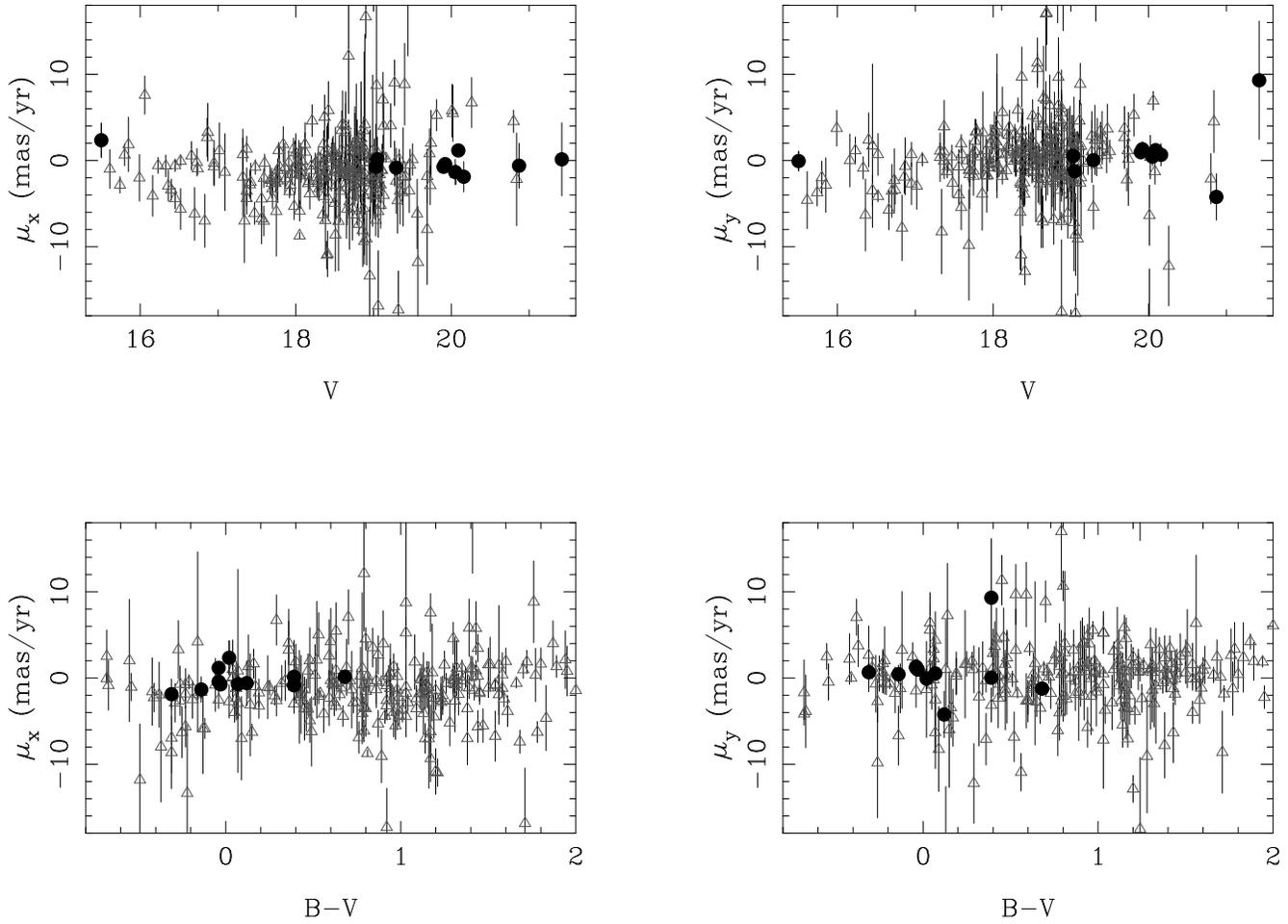}
\caption{Relative proper motions of galaxies and QSOs as a function of
magnitude and color, after the geometric
local correction was applied (see text).}
\end{figure*}

\begin{figure*}
\epsscale{1.1}
\plotone{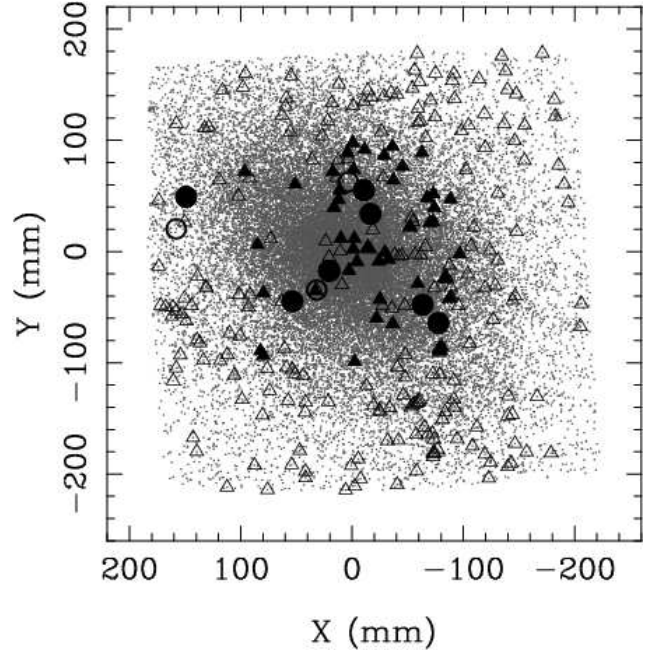}
\caption{Spatial extent of the galaxies and QSOs selected to be used 
in the final absolute proper-motion determination (filled symbols;
circles for QSOs, triangles for galaxies). The open symbols are for 
objects discarded from the final determination (see text).}
\end{figure*}

\begin{figure*}
\epsscale{1.1}
\plotone{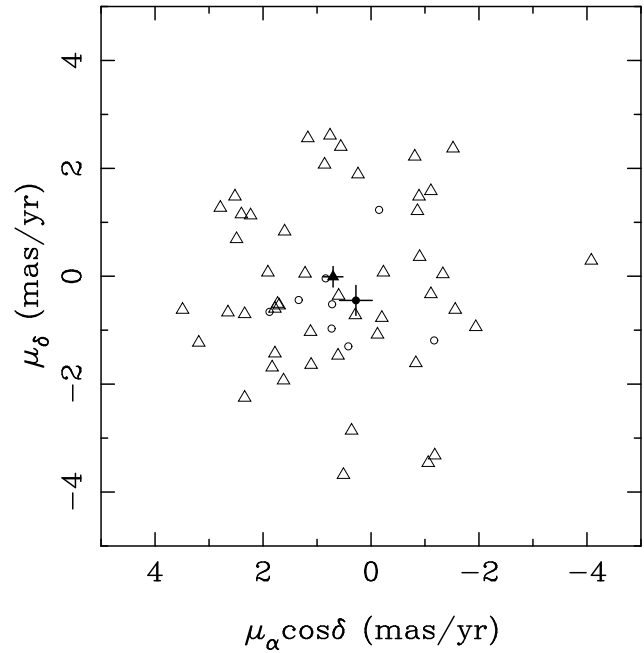}
\caption{Proper motions of the selected QSOs (open circles) and 
galaxies (open triangles). 
The weighted average for the QSOs and the galaxies are indicated with filled 
symbols.}
\end{figure*}

\begin{figure*}
\epsscale{1.8}
\plotone{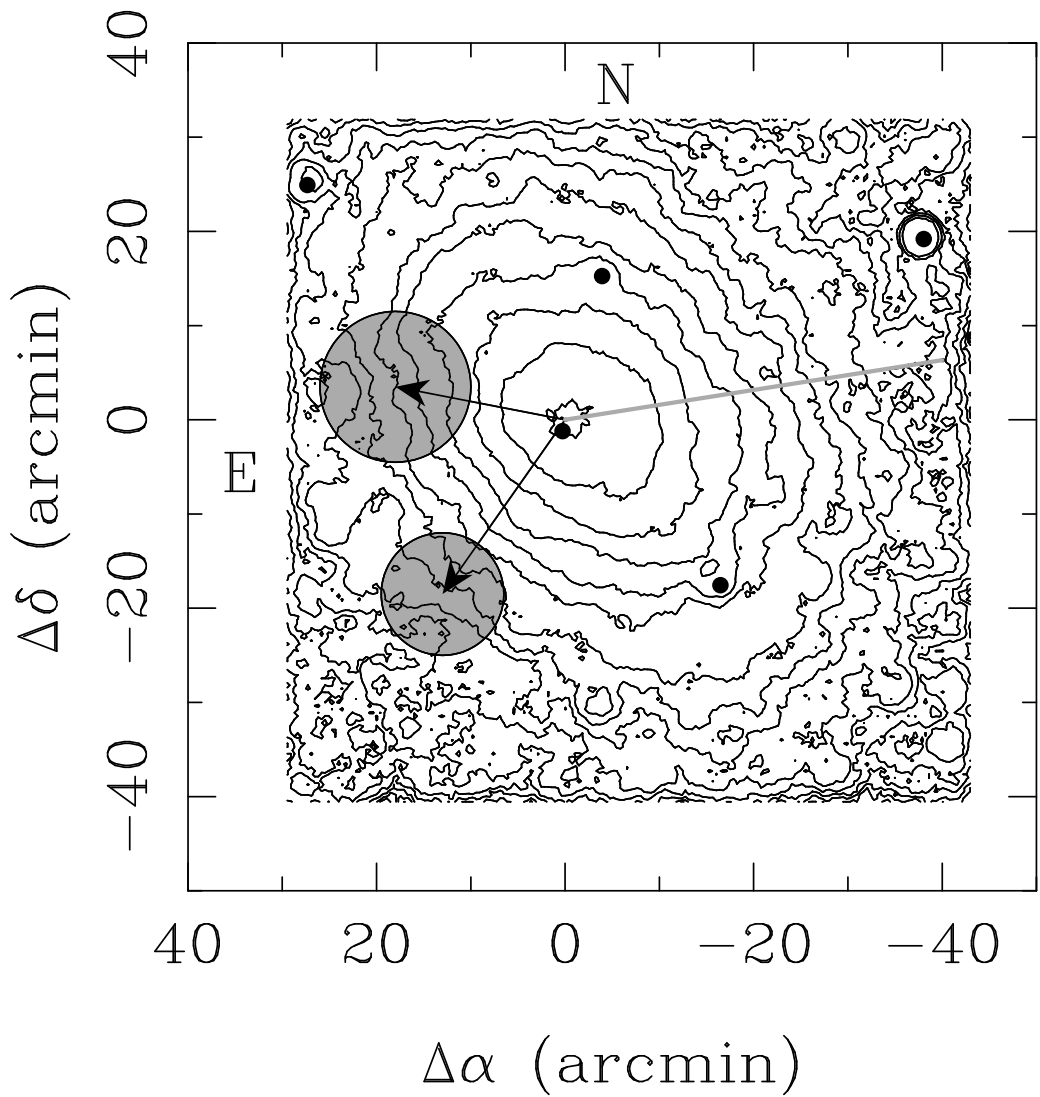}
\caption{The Galactic-rest-frame proper motion of Fornax
superposed on the density contours of Fornax. The short arrow
pointing ENE is this study's determination. The long arrow 
pointing SE is the P02 determination. 
Error circles corresponding to $1\sigma$ uncertainties are indicated.
The gray line shows the direction to Sculptor dSph.}
\end{figure*}

\begin{figure*}
\epsscale{1.3}
\plotone{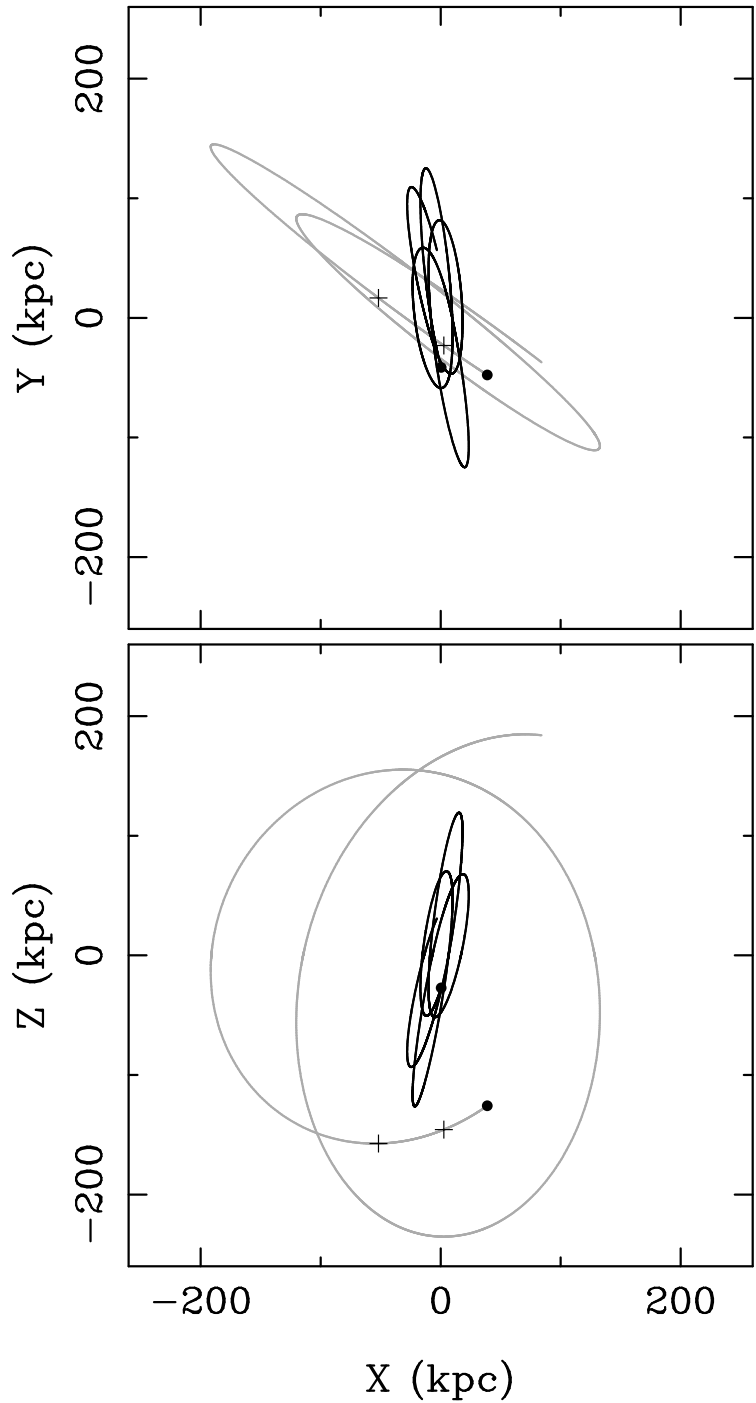}
\caption{The orbits of Fornax (gray line) and the LMC (dark line) integrated back in time for 10 Gyr. The current locations of the LMC and Fornax are indicated with filled symbols. The left panel show the projection on to the Galactic plane, and the right panel that perpendicular to the Galactic plane. The two cross symbols on the orbit of Fornax mark 200 and 500 Myrs ago.}
\end{figure*}

\begin{figure*}
\epsscale{1.4}
\includegraphics[angle=270]{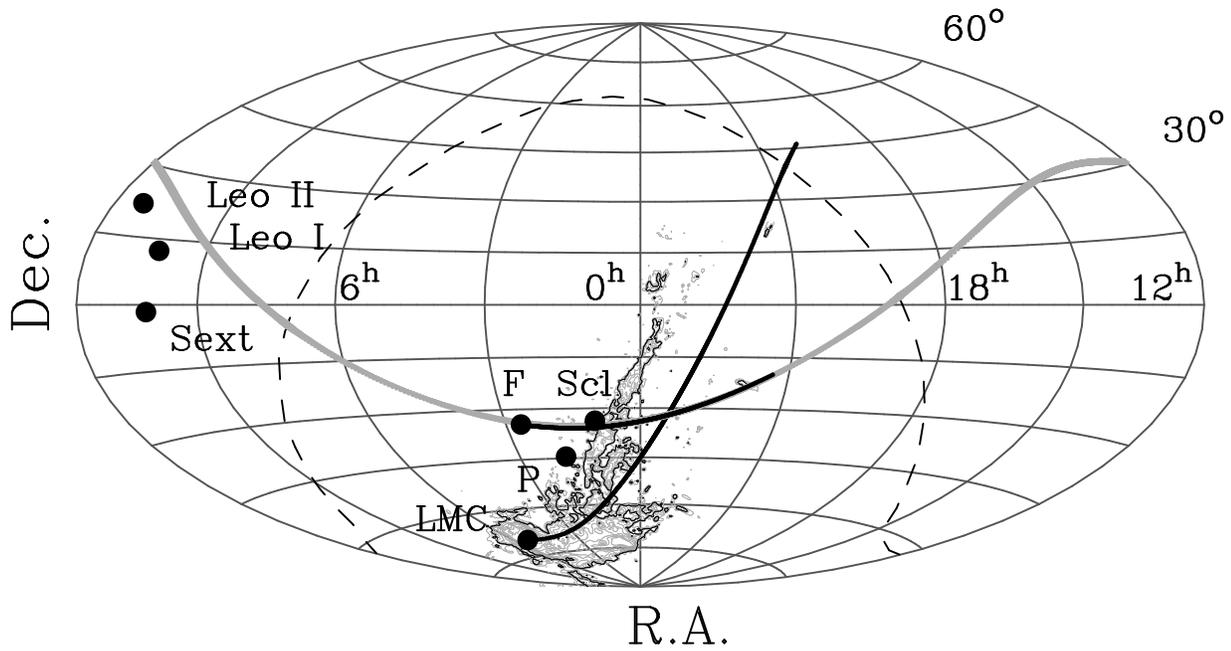}
\caption{Aitoff projection of the orbit of Fornax (gray line).
The superposed black line represents the orbital path in the most recent Gyr.
The most recent Gyr of the orbit of the LMC is also represented with a 
black line. The current positions of the satellites discussed in the text
are represented with filled circles. The dashed line represents the Galactic
plane. The Magellanic Stream is represented with 
H I column density contours (Putman et al. 2003 data). 
The N$_{HI} = 10^{19}$ cm$^{-2}$ level is highlighted.}
\end{figure*}

\begin{figure*}
\includegraphics[angle=270,scale=0.70]{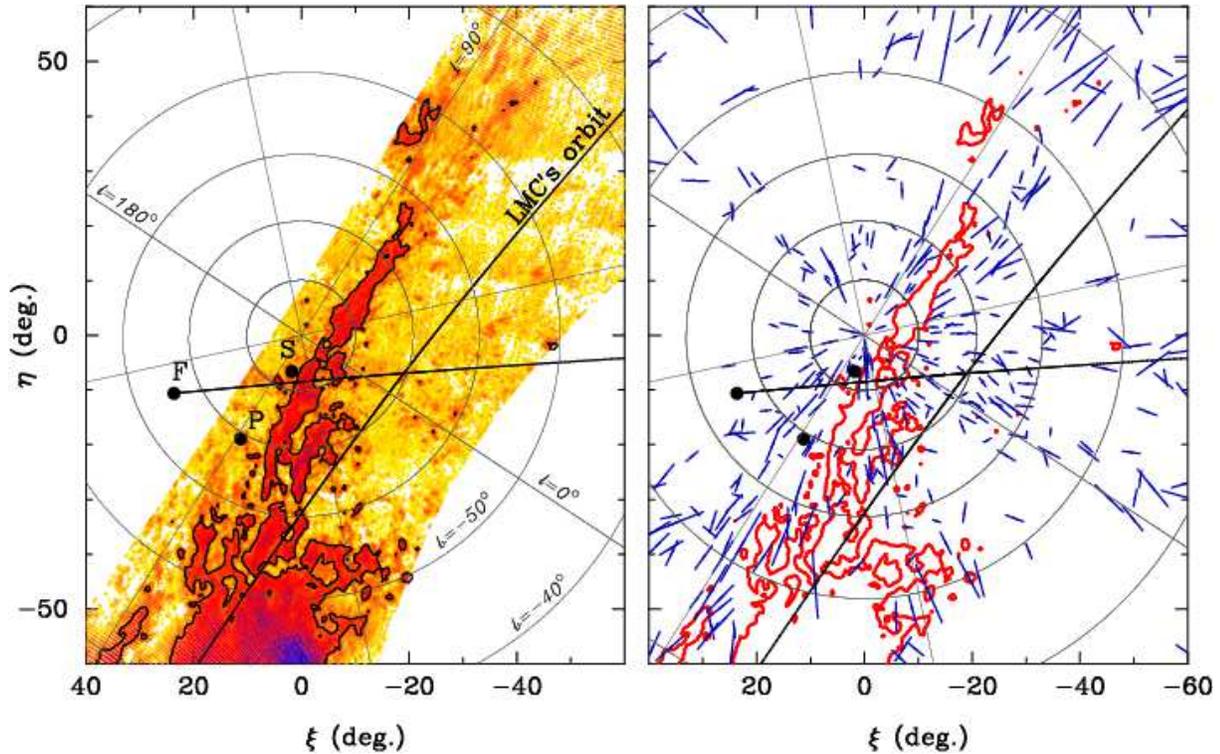}
\caption{The region of the SGP in a gnomonic projection centered on SGP.
The Fornax, Sculptor and Phoenix dSPhs are represented with filled circles,
and are labeled. Fornax's and the LMC's orbit are shown with dark lines.
The left panel (a) shows the column density distribution of H I in the MS
as determined by P03. 
The right panel (b) shows
the distribution of elongated (minor/major axis $\le 0.7$) high velocity
clouds cataloged by Putman et al. 2002. The clouds are represented with
line segments that indicate the size of the major axis, and it orientation.
The N$_{HI} = 10^{19}$ cm$^{-2}$ level in the MS is highlighted in both panels.}
\end{figure*}
\newpage

\end{document}